\documentclass[aps,pre,twocolumn,superscriptaddress,showpacs]{revtex4}
\usepackage{amssymb}
\usepackage{epsfig}
\usepackage{amsmath}
\usepackage{times}

\begin{document}

\title{Variability of Contact Process in Complex Networks}

\author{Kai Gong}
\affiliation{Web Sciences Center, University of Electronic Science
and Technology of China, Chengdu 610054, People's Republic China}

\author{Ming Tang}
\email{tangminghuang521@hotmail.com} \affiliation{Web Sciences
Center, University of Electronic Science and Technology of China,
Chengdu 610054, People's Republic China} \affiliation{Computer
Experimental Teaching Center, University of Electronic Science and
Technology of China, Chengdu 610054, People's Republic of China}

\author{Hui Yang}
\affiliation{Web Sciences Center, University of Electronic Science
and Technology of China, Chengdu 610054, People's Republic China}

\author{Mingsheng Shang}
\affiliation{Web Sciences Center, University of Electronic Science
and Technology of China, Chengdu 610054, People's Republic China}

\date{\today}

\begin{abstract}
We study numerically how the structures of distinct networks
influence the epidemic dynamics in contact process. We first find
that the variability difference between homogeneous and
heterogeneous networks is very narrow, although the heterogeneous
structures can induce the lighter prevalence. Contrary to
non-community networks, strong community structures can cause the
secondary outbreak of prevalence and two peaks of variability
appeared. Especially in the local community, the extraordinarily
large variability in early stage of the outbreak makes the
prediction of epidemic spreading hard. Importantly, the bridgeness
plays a significant role in the predictability, meaning the
further distance of the initial seed to the bridgeness, the less
accurate the predictability is. Also, we investigate the effect of
different disease reaction mechanisms on variability, and find
that the different reaction mechanisms will result in the distinct
variabilities at the end of epidemic spreading.

\end{abstract}

 \maketitle

\textbf{The variability of outbreaks is defined as the relative
variation of the prevalence. In order to assess the accuracy and
the forecasting capabilities of numerical models, the variability
of outbreaks has been investigated in many studies. In numerical
models, many factors such as network structures, travel flows, and
initial conditions can affect the reliability of the epidemic
spreading forecast. Recently, a contact process model with
identical infectivity is proposed to study both dynamical
processes and phase transitions of epidemic spreading in complex
networks, but the predictability of the model is totally
overlooked. In this paper, by investigating the variabilities of
contact process in distinct networks, we show numerically that the
bridgeness plays a significant role on the predictability of the
epidemic pattern in community network, meaning the further
distance of the initial seed to the bridgeness, the less accurate
the predictability is. Hopefully, this work will provide us
further understanding and new perspective in the variability of
contact process in complex networks. }

\section{Introduction}

The great threat of epidemic spreading to human society has been
strongly catching scientists' eyes~\cite{Bailey:1975,
Anderson:1992}. In order to realize the impact of diseases and
develop effective strategies for their control and containment,
the accurate mathematical models of epidemic spreading are the
basic conceptual tools~\cite{Bailey:1975, Anderson:1992,
Diekmann:2000, Dailey:2001}. In mathematical models, the dynamical
patterns of epidemic spreading will be influenced by many
different factors such as the age and social structure of the
population, the contact network among individuals, and the
meta-population characteristics~\cite{Ferguson:2003}. Especially
the heterogeneity of the population network~\cite{Albert:2000} can
result in the absence of endemic threshold when the population
size is infinite and the exponent of degree distribution $\gamma
\leq
3$~\cite{Cohen:2000,Satorras:2001a,Satorras:2001b,May:2001,Lloyd:2001}.
With the further study, the local structures of complex networks
(such as degree correlation, clustering coefficient, community
structure and so on) bring quantitative influences on epidemic
spreading~\cite{Eguiluz:2002,Boguna:2002,Boguna:2003}. Considering
the complicated local structures in real networks, the forecasting
capabilities (i.e. variability) of current numerical models have
been investigated~\cite{Barthelemy:2005}. In addition, both the
stochastic nature of travel flows~\cite{Tang:2006,Kishore:2011}
and initial conditions can affect the reliability of the epidemic
spreading
forecast~\cite{Colizza:2006,Crepey:2006,Gautreau:2007,Gautreau:2008,Tang:2009a,Tang:2009b,Barthelemy:2010}.

In view of this point, Colizza \emph{et al.} have studied the
effect of the airline transportation network on the predictability
of the epidemic pattern by means of the normalized entropy
function~\cite{Colizza:2006}, and found that the heterogeneous
distribution of this network contributes to enhancing the
predictability. In complex networks, many factors can decrease the
forecasting accuracy of epidemic spreading. Cr\'{e}pey \emph{et
al.} have found that initial conditions such as the degree
heterogeneity of the seed show a large variability on the
prediction of the epidemic prevalence, and the infection time of
nodes have non-negligible fluctuations caused by the further
distance and the multiplicity of paths to the
seed~\cite{Crepey:2006}. Comparing the scale-free network (SFN)
with community
structure~\cite{Liu:2005,Huang:2006,Zhou:2007,Liu:2008,Chu:2009,Chen:2009}
with the random SFN, the predictability of the prevalence can be
found to be better~\cite{Huang:2007}.

The common assumption in all the aforementioned works is that each
node's potential infection-activity (infectivity), measured by its
possibly maximal contribution to the propagation process within
one time step, is strictly equal to its degree. However, there are
still many real spreading processes which can not be described
well by this assumption~\cite{Zhou:2006}. Therefore, a contact
process (CP) model with identical infectivity is proposed to study
the epidemic spreading in complex networks~\cite{Castellano:2006}.
Almost all studies in CP are focused on dynamical processes and
phase
transitions~\cite{Zhou:2006,Castellano:2006,Ha:2007,Castellano:2007,
Hong:2007,Yang:2007,Yang:2008a,Yang:2008b,Noh:2009,Lee:2009,Munoz:2010},
but the predictability of the model is totally overlooked. To this
end, we study how the structures of distinct networks (i.e.
homogeneous, heterogeneous and community networks) influence the
variabilities of epidemic patterns in CP. Through numerical
experiments, we find that the community structures can remarkably
influence the prevalence and its variability, contrary to
non-community networks (i.e. homogeneous and heterogeneous
networks). It is worth noting that it's hard for the
extraordinarily large variability in a local community to predict
the epidemic prevalence.

This paper is organized as follows. In Sec. II, we briefly
describe disease models in CP in complex networks and provide
quantitative measurements of the predictability of epidemic
spreading. In Sec. III, we investigate the prevalence
variabilities in both random graph (RG)~\cite{Erdos:1960} and
SFN~\cite{Barabasi:1999}. In Sec. IV, we discuss the essential
differences of the prevalence variabilities both in the global
network and the local community. Finally, we draw conclusions in
Sec. V.

\section{CP model in complex networks}

In our model, three distinct networks, i.e. the homogeneous,
heterogeneous and community networks are adopted to investigate
the predictability of epidemic spreading therein. Firstly, as the
mother of all network models, the random graph of Erd\H{o}s and
R\'{e}nyi~\cite{Erdos:1960} is regularly used in the study of
complex networks because networks with a complex topology and
unknown organizing principles often appear
randomly~\cite{Albert:2002}. Random graph is defined as a graph
with $N$ nodes and connection probability $p$, which has a Poisson
distribution. Secondly, since scale-free property is observed in
many real complex systems, dynamics study on scale-free networks
have been holding everyone's concern. In 1999, Barab\'{a}si and
Albert (BA) put forward the most classical SFN model which is
rooted in two generic mechanisms: growth and preferential
attachment~\cite{Barabasi:1999}. As there are community structures
in social networks, the last studied structure substrate is
community network~\cite{Newman:2002}. Here we will adopt a
simplified community model proposed by Liu and Hu, which
emphasizes on the community feature in social
networks~\cite{Liu:2005}. For simplicity, two independent random
graphs are first produced, and then two RGs are connected randomly
by only one link.

In general, the standard disease models conclude
susceptible-infected (SI), susceptible-infected-susceptible (SIS),
and susceptible-infected-refractory (SIR) epidemiological model.
Each node of the network represents an individual and each link
plays as one connection which transmits disease to other node. In
SI (SIS or SIR) model, 'S', 'I' and 'R' represents respectively
the susceptible (healthy), the infected, and the refractory
(recovered) state. At each time step of contact process, each
infected node randomly contacts one of its neighbors, and then the
contacted neighboring node will be infected with probability
$\lambda$ if it is in the healthy state, or else its state will
stay the same. At the same time, each infected nodes is cured and
becomes susceptible (refractory) with rate $\mu$ in SIS (SIR)
model. To eliminate the stochastic effect of the disease
transmission, we can set $\lambda=1$ and $\mu=0.2$.

In order to analyze the effect of the underlying network topology
on the predictability of epidemic spreading,  the variability of
outbreaks is defined as the relative variation of the prevalence
[density of infected individuals $i(t)$] given
by~\cite{Crepey:2006}

\begin{equation}\label{eq:variability}
\bigtriangleup[i(t)]=\frac{\sqrt{\langle i(t)^{2}\rangle-{\langle
i(t)\rangle}^{2}}}{\langle i(t)\rangle}.
\end{equation}
$\bigtriangleup[i(t)]=0$ denotes all independent dynamics
realizations are essentially the same, and the prevalence in the
network is deterministic. Larger $\bigtriangleup[i(t)]$ means
worse predictability that a particular realization is far from
average over independent realizations.

\section{Predictability in homogeneous and heterogeneous networks}

The first issue of our study is how the heterogeneity of network
structures influences the variability of the prevalence in CP. By
using a numerical approach in this section, we analyze the
variabilities of outbreaks generated by different sets of initial
nodes, both for random graphs and scale-free networks with the
same network size and average degree. Considering the fact that
the results of a particular network can be generalized to any
instances of network model~\cite{Crepey:2006}, the numerical
simulations we studied here are run in one network. In Fig. 1, we
show the curves $i(t)$ and $\bigtriangleup[i(t)]$ computed for the
different disease models in both RG and SFN. For SI model in Fig.
1 (a), the density of infected $i(t)$ in RG reachs its stationary
state faster than that in SFN; for SIS model in Fig. 1 (b), the
stationary $i(t)$ in RG is greater than that in SFN; and for SIR
model in Fig. 1 (c), RG has the higher peak prevalence. Contrary
to the results for the case of contacting all neighbors, it is
first discovered that the heterogeneous structure can slow down
the prevalence of outbreaks in CP. Because hubs may be contacted
many times by their neighboring nodes at each time step, the total
contact ability (i.e. the actual number of contacting nodes at one
time step) of SFN is reduced further accordingly, as a result, the
hub effect holds back the prevalence of diseases. Meaningfully,
owing to the limited contact ability of CP, the infected densities
starting from the initial infected nodes (seeds) with different
degrees are almost the same in SFN, which is distinct from the
results for the case of contacting all
neighbors~\cite{Crepey:2006}.

\begin{figure}
\epsfig{figure=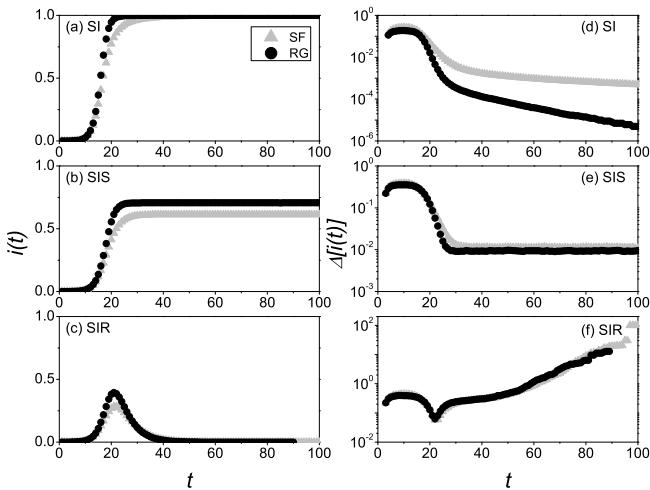,width=1.0\linewidth}\caption{(color
online). Evolution of both $i(t)$ and $\bigtriangleup[i(t)]$ for
the different disease models where the "triangles" and "circles"
denote the cases of SF and RG networks with the random initial
seeds. $i(t)$ versus $t$ for SI model (a), SIS model (b), and SIR
model (c), and $\bigtriangleup[i(t)]$ versus $t$ for SI model (d),
SIS model (e), and SIR model (f). The parameters are chosen as
$N=0.5\times10^{4}, \langle k\rangle=10, \lambda=1$, and
$\mu=0.2$. The results are averaged over $2\times10^{4}$
independent realizations in one network.} \label{Fig:BA_ER}
\end{figure}

As shown in Fig.~\ref{Fig:BA_ER} (d), (e), and (f), there is
slightly different between variabilities in RG and SFN when
$t<20$, which implies heterogeneous structure does not visibly
alter the predictability of CP before the outbreak of disease. An
important contribution of this study is to analyze the differences
among the variabilities of three kinds of disease models. From the
comparison among them, we find that different recovery mechanisms
can result in distinct variabilities at the end of epidemic
spreading. For SI model in Fig. 1 (d), the time arriving at
$i(t)=1$ varies in a mass of realizations, which can induce an
exponential decay of the variability when $t>30$. For SIS model in
Fig. 1 (e), the variability of the prevalence will keep on a
steady value in stationary state. For SIR model in Fig. 1 (f), due
to the different lifetimes of the epidemics in a mass of
independent realizations~\cite{Colizza:2006}, the greater and
greater variabilities are observed by approaching the end of the
epidemics.

% figure 1

\section{predictability in the global network and the local community}

\subsection{Global network}

% figure 2
\begin{figure}
\epsfig{figure=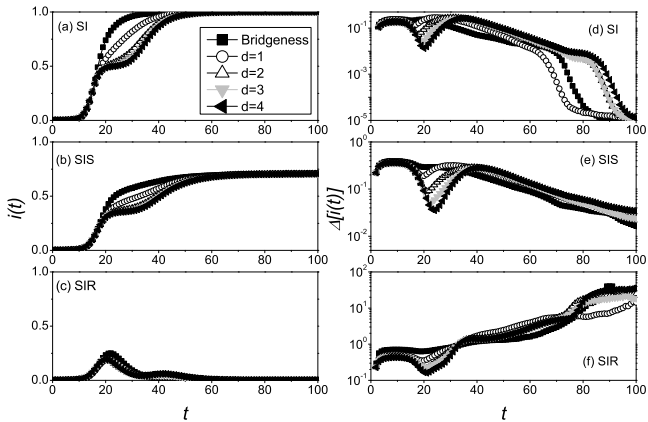,width=1.02\linewidth}\caption{(color
online). Evolution of both $i(t)$ and $\bigtriangleup[i(t)]$ in
community networks where the "squares", "circles", "triangleups",
"triangledowns", and "trianglelefts" denote the cases of the
bridgeness, $d=1,2,3$, and $4$, respectively. $i(t)$ versus $t$
for SI model (a), SIS model (b), and SIR model (c), and
$\bigtriangleup[i(t)]$ versus $t$ for SI model (d), SIS model (e),
and SIR model (f). The parameters are chosen as $N=10^{4}, \langle
k\rangle=10, \lambda=1$, and $\mu=0.2$. The results are averaged
over $2\times10^{4}$ independent realizations.}
\label{Fig:Network}
\end{figure}

As many social networks combined by several communities, such as
Facebook~\cite{Facebook}, YouTube~\cite{Youtube}, and
Xiaonei~\cite{Xiaonei}, information propagation taking place in
community
networks~\cite{Liu:2005,Huang:2006,Zhou:2007,Liu:2008,Chu:2009} is
one of the most important subjects studying in complex networks,
but in CP, the related research has been ignored for a long time.
Therefore, in this section, we study the variability of CP in a
very simple community network, where two RGs are connected
randomly by only one link. Obviously, this network has a strong
strength of community structure. In order to normalize the terms
of community network, we define the link as weak
tie~\cite{Onnela:2007}, and call two nodes connected by this link
"bridgeness"~\cite{Cheng:2010}. We first investigate the time
evolution of epidemics generated by different seeds staying away
from the bridgeness, and it is noted that there is only one
initial seed in each realization. From Fig. 2 (a), (b), and (c),
we can get that the closer the seed to the bridgeness,the epidemic
spreads much faster in the global network, and among all cases,
the epidemic starting at the bridgeness spreads fastest. For SI
model, the further distance to the bridgeness such as $d=3, 4$
induces two periods of the quickly rising trend at the beginning
time $t=10,30$, respectively. If the initial seed is far away from
the bridgeness, the disease will be restricted in the first
community for a long time till the bridgeness infected, in which
almost all nodes is infected. As a result, the outbreak in the
second community just starts at that moment the prevalence in the
first community get towards the end, which causes the second
outbreak. In the case of SIS and SIR model, the recovery mechanism
reduces this phenomenon occurred, for instance there is the tiny
second peak of the prevalence for $d=3,4$ in Fig. 2 (c). From the
above, it is found that the bridgeness plays a distinctly
important role in the rapid transmission of information in CP.

The variability of prevalence in community network is distinct
from that in the network which has no community structure. As
shown in Fig.~\ref{Fig:Network}(d), (e), and (f), the curves
$\bigtriangleup[i(t)]$ display two peaks because of the time delay
between two outbreaks occurred in different communities. In
addition, the further distance to the bridgeness makes the second
peak occur much later. In Fig.~\ref{Fig:Network}(d), for SI model,
the first peak corresponds to the prevalence in the community with
the initial seed, so the variability is almost the same as that in
RG before $t=10$. With the outbreak in the second community, the
second peak occurs. Owing to the greater randomness of the time
that disease first occurs in the second community (see
Fig.~\ref{Fig:arrival time}), the second peak of the variabilities
is slightly greater than the first peak. After the infection
density is close to saturated at $t\approx40$ (see
Fig.~\ref{Fig:Network}(a)), the variability will be on exponential
decay. As all nodes of community network are infected in more and
more realizations, the variability $\bigtriangleup[i(t)]$ will
rapidly decay to zero. In contrast with the case of SI model, the
second peak in SIS model is less than the first peak, because the
recovery mechanism slows down the propagation velocity of
diseases, which reduces the variability of the prevalence. That is
to say, the recovery mechanism reduces the variability of epidemic
spreading. Another extremely obvious difference is the variability
decreases to a steady value at the stationary state. For SIR model
in Fig.~\ref{Fig:Network}(f), as time goes by, the epidemics has
the greater and greater variability, which is caused by the
different lifetimes of the epidemics in $2\times10^{4}$
independent realizations.

\subsection{The local community}

% figure 3
\begin{figure}
\epsfig{figure=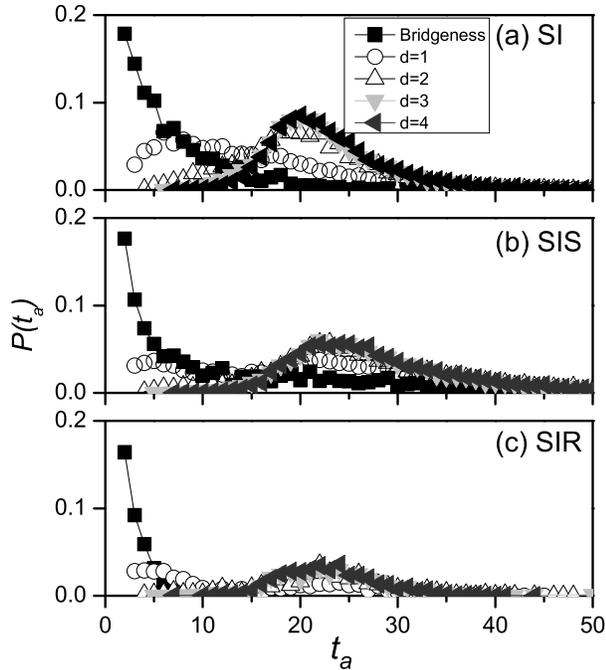,width=1.0\linewidth}\caption{(color
online). The distribution of arrival time of disease in the second
community for SI model (a), SIS model (b), and SIR model (c),
where the "squares", "circles", "triangleups", "triangledowns",
and "trianglelefts" denote the cases of the bridgeness, $d=1,2,3$,
and $4$, respectively. The results are averaged over
$2\times10^{4}$ independent realizations.} \label{Fig:arrival
time}
\end{figure}

% figure 4
\begin{figure}
\epsfig{figure=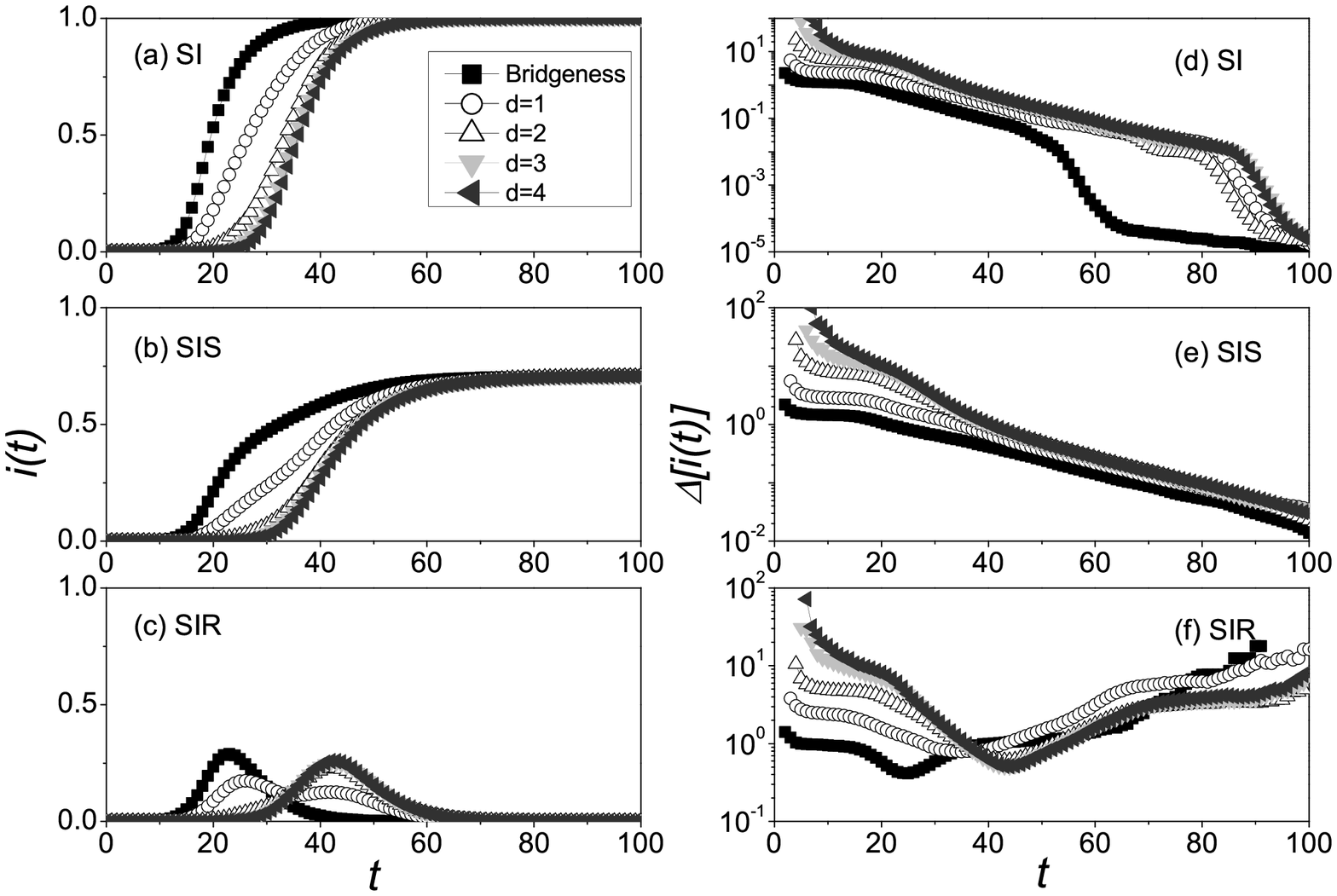,width=1.05\linewidth}\caption{(color
online). Evolution of both $i(t)$ and $\bigtriangleup[i(t)]$ in
the second community where the "squares", "circles",
"triangleups", "triangledowns", and "trianglelefts" denote the
cases of the bridgeness, $d=1,2,3$, and $4$, respectively. $i(t)$
versus $t$ for SI model (a), SIS model (b), and SIR model (c), and
$\bigtriangleup[i(t)]$ versus $t$ for SI model (d), SIS model (e),
and SIR model (f). The results are averaged over $2\times10^{4}$
independent realizations.} \label{Fig:Community}
\end{figure}

Considering the relative independence of a local community, we
should take the prevalence and variability into account. On the
other hand, since disease must be transmitted through bridgenesses
from the first community to the second community, this study
contributes to understand the effect of them on epidemic
spreading~\cite{Zhao:2010}. In this section, we will specifically
analyze the effect of different distances of seeds (to the
bridgeness in the first community) on epidemic spreading in the
second community.

At first, the arrival time of disease is defined as the moment
that infectious individual first occurs in the second community in
each realization, thus the distribution of arrival time is
obtained through massive realizations. In Fig. 3, the distribution
of arrival time for the different initial seeds is showen. For SI
model, the arrival distribution of the bridgeness as seed $d=0$
strictly obey the distribution $P(t)=\lambda(1-\lambda)^{t-1}$.
When disease seed is the node with one step to bridgeness, the
arrival time increases generally, and the distribution becomes
much wider and flatter. With the further increasing of distance of
seed to the bridgeness (such as $d=3,4$), the distributions are
nearly the same. It is understood that due to the finite size
effect of network, the disease is transmitted through weak tie to
the second community till overall outbreak happened in the first
community. For SIS model, as a result of the recovery mechanism,
the distributions of arrival time are much more evenly and
smoothly than that for SI model, given the various initial seeds.
Compared with Fig.~\ref{Fig:arrival time} (a), we can find that
there are two peaks for SIS and SIR model with $d=1$. This is
because the bridgeness may be infected through two basic pathways:
the bridgeness may be infected directly by the initial seed (i.e.
its neighboring node) in $t\leq1/\mu$; the other route is the
transmission of infection from the other neighboring nodes when
the disease outbreak in the first community, thus the second peak
occurs at $t\approx20$ (see Fig.~\ref{Fig:Network}(b)). For SIR
model, owing to the recovery mechanism, the prevalence might well
disappear in the first community before arriving at the weak tie.
Consequently, arrival rate (i.e. the area of the distribution) is
less than one, and the peak value of the corresponding
distribution is less than that for both SI and SIS model.

Fig.~\ref{Fig:Community} shows the prevalence and variability in
the second community which are generated from the different
initial seeds in the first community. From
Fig.~\ref{Fig:Community} (a), (b), and (c),we can easily know that
the prevalence of the bridgeness acting as seed increases much
faster than that in the other cases generated by the further
seeds. In particular, for SIR model in Fig.~\ref{Fig:Community}
(c), the peak of the prevalence (for the case of the bridgeness)
occurs first, and has the maximum value. In addition, there are
two peaks of the prevalence for the case of $d=1$, which is
attributed to the propagation delay between two communities. Since
there is only one interconnection between two communities, the
chance of infecting one from the other is low. In other words,
infection within intra-community will be much faster than
inter-community infection. Therefore, the first peak reflects the
outbreak and extinction of disease inside with infectious seed,
and the second peak emerges after the other community is infected.
Note that the two peaks can be only observed in SIR model because
of the fact there will be no peak if virus does not "die". With
the increase of distances $d$, the peak value is greater than that
for $d=1$, although the outbreaks occur later.

In Fig.~\ref{Fig:Community} (d), (e), and (f), it is a surprise
that the variabilities in the second community are distinct from
that in the global network. Firstly, the variabilities in the
second community are very large and also much greater than that in
the global network in Fig.~\ref{Fig:Network}, which implies the
huge unpredictability of prevalence produced in the local
community. In particular, at the beginning of outbreaks, the
variability for the case of $d=4$ reaches about $50$, which is 100
times the maximum value $0.5$ in the global network. Secondly, the
closer distance of the seed to bridgeness, the lower level of
variability it has in the local community. In particular, the
maximum variability value for the case of the bridgeness is only
about $1$, which is much less than $50$ for the case of $d=4$.
Thus, the bridgeness plays a significant role in enhancing the
predictability, that the closer initial seed to the bridgeness,
the more accurate the predictability is. Thirdly, each curve of
variabilities can be divided into four parts: the sudden drop
stage, the relatively stable stage, the slowing-down stage, and
the final stage of outbreaks (i.e. the exponential decay stage for
SI model, the steady state stage for SIS model, and the sharp
increase stage for SIR model, respectively). The first stage is
originated from the uncertain arrival time of disease, with the
increase of the arrival rate, the variability decreases. The
second stage ($10<t<20$ in most cases) is induced by the interplay
between the outbreak and the arrival of diseases in massive
realizations: on the one hand, the outbreaks in some realizations
upgrade the variability; on the other hand, the increase of the
arrival rate counteracts this effect. As the infection density is
close to saturated, the variability will enter the slowing-down
stage. What is noteworthy is that the minimum variability value
just corresponds to the peak value of prevalence for SIR model. In
the end of epidemic spreading, the variabilities
$\bigtriangleup[i(t)]$ display the distinct phenomena for the
different disease models, for instance, SIR model shows the higher
and higher variabilities.

\section{conclusions and discussions}

In conclusions, we have studied the variability of CP in complex
networks, and get the clear understanding that the different
network structures can remarkably influence the prevalence and its
variability. Firstly, we find that the variability difference
between homogeneous and heterogeneous networks is very narrow,
although the heterogeneous structure induces a lighter prevalence.
Secondly, two peaks of both the prevalence and variability are
shown in the community network. It's noted that in the local
community, the extraordinarily large variability in early stage of
the outbreak makes the prediction of disease spreading hard. This
result is in accordance with Ref.~\cite{Yan:2007} in which the
networks with strong community structures are of weak
synchronizability, and the amplitudes of the time series in the
local communities are much larger than that in the global
networks. Fortunately, the bridgeness plays a significant role in
enhancing the predictability, the closer initial seed to the
bridgeness, the more accurate the predictability is. This result
suggests that bridgenesses may be the ideal detection stations in
community networks. Moreover, the different reaction mechanisms of
disease models can result in the distinct variabilities.
Especially for the case of SIR model, the greater and greater
variabilities are observed at the end of the epidemics for the
different lifetimes of the epidemics in various realizations.

The community network employed in this study is much more simple,
but the actual community networks have complex structures, such as
multifarious communities, many bridgenesses, and the heterogeneous
degree distribution in a local community. Therefore, the further
investigation should be focused on the more complex community
networks.

\acknowledgments This work is supported by the NNSF of China under
Grant Nos. 90924011, and the Sichuan Provincial Science and
Technology Department (Grant No. 2010HH0002).

\end{document}